\documentclass[twocolumn,twocolappendix]{aastex63}

\shorttitle{Raman scattering in the general interstellar medium}
\shortauthors{F. Zagury}

\begin{document}

\title{Raman Scattering  as the Key Link Between Unidentified Infrared Bands, Diffuse Interstellar Bands, and Extended Red Emission}

\correspondingauthor{Fr\'ed\'eric Zagury}
\email{fzagury@wanadoo.fr}

\author[0000-0002-0786-7307]{Fr\'ed\'eric Zagury}
\affiliation{Fondation Louis de Broglie \\
23 rue Marsoulan \\
75012 Paris, France}




\begin{abstract}
This paper calls attention to the relevance of Raman scattering by atomic hydrogen to three optical and near/mid-infrared spectral features of HI clouds: extended red emission (ERE), diffuse interstellar bands (DIBs), and the unidentified infrared bands (UIBs). DIBs, ERE, and UIBs are observed predominantly at the edge of HI clouds, are manifestly related, and remain poorly understood. Their salient properties correspond to two major characteristics of HI Raman scattering: unusual  line broadenings and a concentration of the Raman scattered ultraviolet continuum in the vicinity of hydrogen’s optical and infrared transitions. Raman scattering by atomic hydrogen has now been detected in several object classes where the spectral features are observed, and I argue that it can account for all three features.
I further identify three factors that condition observation of Raman scattering in HI clouds, and thus of DIBs, ERE, and UIBs: the hardness of the radiation field, interstellar dust extinction, and the geometry of the observation. The geometry determines whether complete forward scattering, yielding DIBs, or scattering at large angles, yielding ERE in the vicinity of $H\alpha$ and UIBs in the infrared spectrum, will be observed. ERE results from Raman scattering of photons near $Ly\beta$ and UIBs from excitation of hydrogen atoms close to the ionization limit. DIBs, ERE, UIBs are thus different facets of the same interstellar phenomenon: Raman scattering by atomic hydrogen.
\end{abstract}

\section{Introduction}\label{intro}
\begin{quote}
"We show that Raman scattering deserves to be included in the diagnostic tools of the spectroscopy of gaseous nebulae and emission regions of galactic nuclei. We point out that Raman scattering may be the source of some up to now unidentified emission lines."
\citep[][p.~L27]{nuss89}.
\end{quote}

The quoted passage refers to Raman scattering by atomic hydrogen, and  the term "gaseous nebulae" should include all HI interstellar media illuminated by an ultraviolet radiation field. 
In HI media an observed optical or infrared emission or absorption line may be the "ghost"  of a true emission (or absorption) ultraviolet line Raman scattered by hydrogen.
The observed line will be broadened by  a factor of 40 ($\equiv \,(H\alpha/Ly\beta)^2$, Section~\ref{rs}) close to $H\alpha$, and at least a hundred times more if the ghost is in the mid-infrared.
It follows that broad unidentified emissions or absorptions from HI regions raise the suspicion that Raman scattering is at work.

The discovery of Raman scattering by atomic hydrogen in interstellar space was fairly recent.
In 1989  H.M.~Schmid   showed how Raman scattering could account for the puzzling width of a pair of optical lines that seemed too large for a traditional Doppler effect \citep[][]{schmid89}. 
Since  he could derive the  positions at which the lines were observed and their exact ratio, his interpretation in terms of Raman scattering was immediately accepted.
Together with the article by  \cite{nuss89} his discovery stimulated a new line of research.
Emission lines  Raman scattered by hydrogen  have since been found in planetary nebulae \citep[PNe,][]{pequignot97,lee06} and galaxies  \citep[][]{dopita16}.
Scholars have also been interested in Raman scattering of the ultraviolet continuum  \citep[][]{lee00a,jung04,dopita16}.
The work of  \citet[][]{chang15} suggests that hydrogen Raman scattering of the ultraviolet continuum  clusters around hydrogen transitions, with strong minima in between.
\citet{henney21} recently detected a pair of oxygen Raman scattered lines, in absorption this time, from photodissociation regions of the Orion bar that  Raman scatter the ultraviolet continuum of Orion stars.

Curiously enough, no connection has been made between these investigations of hydrogen Raman scattering and major problems that occupy research on the optical and infrared properties of interstellar clouds (extended red emission, diffuse interstellar bands, and unidentified infrared bands).
Yet they concern the same media (HI clouds), and even the same objects (for instance NGC7027 or the Orion bar), similar column densities (usually in the range $10^{19}$ -- $10^{22}$~cm$^{-2}$), and  illumination by ultraviolet-dominated radiation fields (color temperatures generally above $\sim10^4$~K).

Extended red emission (ERE) is a wide emission-like bump between 6000 and 8000~\AA\ in the vicinity of $H\alpha$, on top of which lies a series of unidentified broad emission bands.
ERE has been observed in nebulae and galaxies \citep{witt90,perrin95}.
Also in the optical part of the spectrum, diffuse interstellar bands (DIBs) are absorptions exclusively observed in the spectrum of stars in the background of interstellar clouds.
DIBs are often unusually broad \citep[Figure~7 in][]{hobbs09}, and it has been shown that some of the strongest DIBs lie on the blue edge of the main ERE emission lines \citep{vanwinckel02}.
In the infrared part of the spectrum,  researchers are grappling with the series of unidentified infrared bands (UIBs) and their underlying wavy continuum  \citep[Figure~1 in ][]{peeters04}.
In nebulae where both ERE and near-infrared excesses are observed they occupy the same spatial domain \citep{sellgren84,tuthill02}. 
All three features (ERE, DIBs, infrared excesses) are known to peak in HI-rich photodissociation regions \citep[PDRs,][]{boulanger98,herbig95,lai20}.

On the assumption that the observed emission/absorption optical and infrared lines should be the focus of explanation, great efforts have been expended to date on identifying particles that would emit/absorb at these wavelengths.
However,  the hypothesized existence in interstellar space of ”a new component of the interstellar matter” \citep{puget89} consisting of polycyclic aromatic hydrocarbons  \citep[PAHs,][]{leger84}  has thus far yielded no conclusive results  \citep[see Introduction in][]{sorokin00}.
These complex molecules should pervade the Galaxy down to its least dense regions \citep{kahanpaa03}, including  interstellar clouds with so few H-atoms per cm$^3$  that  H$_2$  cannot be detected.
No such molecules have been identified.
PAH theory further requires an improbably large number of independent parameters  \citep{sadjadi15,zhang15,param}.

The broadness of UIBs, of the Red Rectangle's emission lines, and of DIBs, and the location of ERE on the spectrum (around $H\alpha$), are suggestive of Raman scattering by hydrogen.
Consistent with Nussbaumer et al.'s intuition, the reason why these features remain unidentified could be that they originate from wavelengths other than the observed ones.
The question thus arises whether Raman scattering by hydrogen could provide the basis for an alternative  account of ERE, DIBs, and UIBs.
This possibility is strengthened by my recent finding that the imprint of UIBs on the spectrum correlates with hydrogen transitions \citep{uib}.
The whole ERE/UIB spectrum, which is basically the spectrum of a nebula (Section~\ref{patn} and Appendix~\ref{sell}), must therefore overlap with the spectrum of hydrogen.
Further, I will show in this paper that the two aforementioned oxygen pairs of Raman scattered  lines are found in  the latest DIB catalogs (Section~\ref{dibid}).

To explore the hypothesis that Raman scattering can provide a unified understanding of ERE, DIBs, and UIBs is the purpose of this paper.
The first part of the paper (Sections~\ref{rs}--\ref{rsca}) introduces two properties of hydrogen Raman scattering that are relevant for the study of the interstellar medium.
One is that ultraviolet extinction (gas + dust) brings Raman scattering by hydrogen to saturation (Section~\ref{dext}), leading to a specific pattern of  the Raman spectrum  (Section~\ref{patn}). 
The other is that the Raman brightness of a nebula around  $H\alpha$ exceeds its Rayleigh brightness by several orders of magnitude when the color temperature of the illuminating star is above $\sim10^4$~K  (Section~\ref{rsca}).

A second part (Section~\ref{scage}) highlights  the dependence of reflected starlight from HI media on the angle of scattering (viewed from the observer position).
For a radiation field with a color temperature above $\sim10^4$~K,  reflected light from an HI cloud at optical and infrared wavelengths is Raman scattered light of ultraviolet photons at   large scattering angles.
The scattering is inelastic and near-isotopic.
When the angle of scattering diminishes down to near-forward directions, elastic forward scattering by interstellar dust will prevail with  a $1/\lambda^p$ ($p\simeq 1$) extinction law.
At null angle  gas (Raman) scattering becomes coherent and  overwhelms dust scattering.
Each type of scattering is characterized by a different type of spectrum and  determines whether ERE and UIBs or DIBs will be observed.

The  third part of the paper  (Sections~\ref{ere}--\ref{ex}) analyzes observations of ERE (Section~\ref{ere}), DIBs (Sections~\ref{ed}--\ref{dib}), and UIBs (Section~\ref{uib}) in terms of Raman scattering.
Because of the questions they raise, the cases of NGC7027 and IC60 are briefly considered in Section~\ref{ex}.

The conclusion sums up my argument that, notwithstanding continuing enigmas, there is sufficient evidence to think that only a few conditions (an ultraviolet radiation field, dust extinction, the angle of scattering) permit the expression of Raman scattering in interstellar space through the ERE, DIBs, and UIBs, and  that these spectral features are different facets of the same process.
An Appendix expresses my doubts about K.~Sellgren's claim \citep{sellgren83t}  that near-infrared excesses in nebulae cannot be scattered light and must result from thermal emission by transiently heated small particles.
This claim was instrumental in the development of the PAH hypothesis \citep{leger84} and has never been questioned.
But it introduces a modeling of scattering by small interstellar grains that is not supported by observation and rests on misinterpretations of interstellar polarization data.

\section{Raman shifts and line-broadening} \label{rs}
Raman shift 
\begin {equation}
\nu_R=1/\lambda_R=1/\lambda_i-1/\lambda_f , \label{eq:rs}
\end {equation}
is a major characteristic of a Raman scattering process.
It corresponds to the wave-number difference between the initial and final states of the particles that Raman-scatter light and fixes the energy gap between a source photon of wavelength $\lambda_i$ and the Raman scattered photon at wavelength $\lambda_f$.

Molecular Raman scattering shifts do not exceed a few 1000~cm$^{-1}$ \citep[4161~cm$^{-1}$ for the vibrational Stoke $Q(1)$ Raman  transition of molecular hydrogen,][]{cochran78}.
The corresponding wavelength differences between incoming and outgoing photons are moderate ($\sim1000$~\AA\ for near-ultraviolet photons Raman scattered by H$_2$).

The minimum Raman shift for Raman scattering by atomic hydrogen at rest is 82259~cm$^{-1}$ (Section~\ref{data}), twenty times that of molecular hydrogen.
An ultraviolet photon with wavelength $\lambda_i$ less than ${Ly\alpha}$ is  converted into an optical or infrared photon.
To compare  the efficiency of Raman relative to Rayleigh scattering it is therefore necessary to use the  Rayleigh scattering cross-sections at optical and infrared wavelengths, which are orders of magnitude smaller than at ultraviolet wavelengths.
This comparison also  needs to incorporate the relative intensities of the radiation field at the source and scattered wavelengths (Section~\ref{ramray}).
The elastic scattering rule that scattering is optimized for optical depths close to 1 does not hold anymore (Section~\ref{gext}).

Expressed differently, Equation~\ref{eq:rs} relates source photons at wavelengths $\lambda_i$ and $\lambda_i +\Delta \lambda_i$ and Raman-scattered photons  at  $\lambda_f$ and $\lambda_f +\Delta \lambda_f$  by the line-broadening formula   \citep[Equation~7 in][]{nuss89}
\begin {equation}
\frac{\Delta \lambda_f}{\lambda_f}=\frac{\lambda_f}{\lambda_i} \frac{\Delta\lambda_i}{\lambda_i}. \label{eq:br1}
\end {equation}
Rewritten as 
\begin {equation}
\Delta \lambda_f=\left( \frac{\lambda_f}{\lambda_i} \right)^2\Delta\lambda_i, \label{eq:br2}
\end {equation}
the broadening formula  expresses that Stokes Raman scattering by atomic hydrogen converts an ultraviolet wavelength interval into a much wider optical or infrared one. The energy scattered is likewise diluted.
Emission and absorption lines are  broadened.
\section{Ultraviolet extinction in HI clouds} \label{sigma}
\subsection{Data} \label{data}
\begin{table}[b]
\begin{center}
\caption{Characteristic values for Raman scattering by hydrogen at rest and left at levels $n=2$, 3, 4.
Columns~2 to 6 give, for each level, the Raman shift, the maximum wavelength and minimum energy required for source photons (their energy must be greater than transition $T_{n+1\rightarrow n}$), the minimal wavelength and maximum energy of Raman scattered photons (Raman scattering of $Ly_\infty$ photons).
\label{tbl:rcar}}		
\begin{tabular}{crrcrc}
level&$\nu_R$ $\,\,\,\,$&$\lambda_{i,max}$&E$_{i,min}$ &$\lambda_{f,min}$&E$_{f,max}$\\
&cm$^{-1}$   & \AA\ $\,\,\,\,$   &eV& \AA\ $\,\,\,\,$&eV \\
$n=2$& 82258&1215.7&10.2 &3647&  3.40  \\
$n=3$&97491&1025.7&12.1&  8206 & 1.51 \\
$n=4$& 102823&972.5&12.7&1.46 $\mu$m & 0.85  \\
\end{tabular}
\end{center} 
\end{table}
\begin{figure*}[]
\resizebox{2.\columnwidth}{!}{\includegraphics{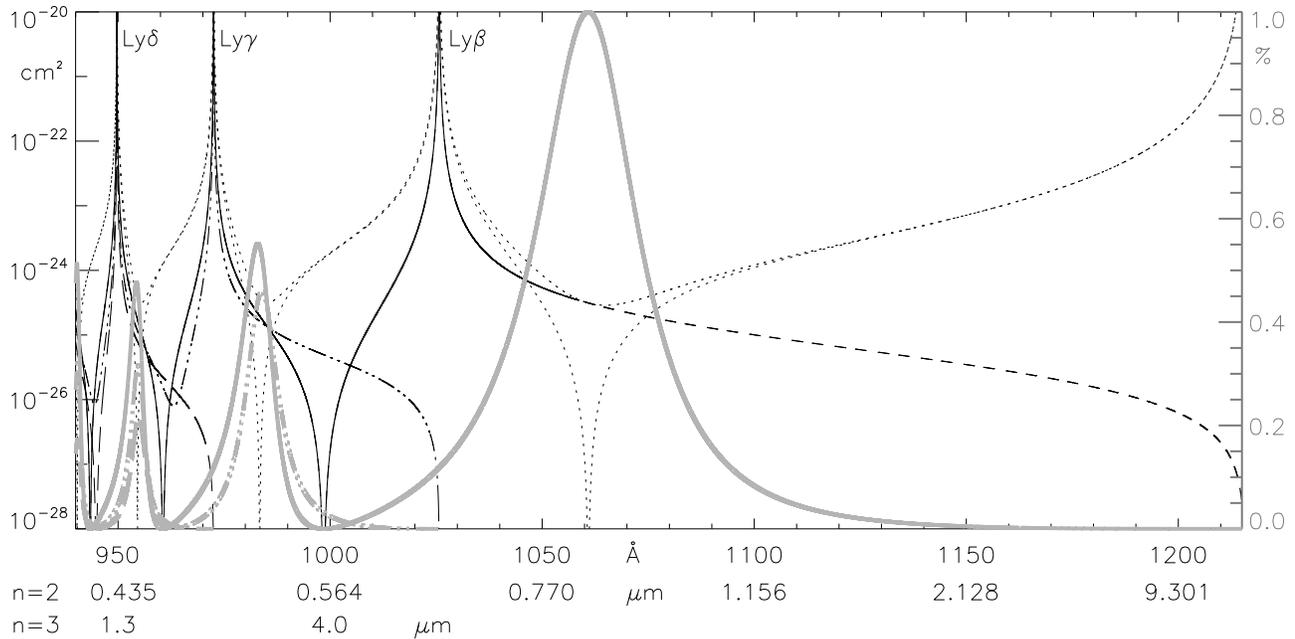}} 
\caption{Atomic hydrogen scattering parameters in the $Ly\epsilon$--$Ly\alpha$ wavelength range. 
Total (Raman +Rayleigh) and Rayleigh cross-sections are both in dots.
Raman cross-sections for $n=2$, 3, 4 are the solid (small dashes for wavelengths imaged above 8000~\AA), dotted-dashed, and dashed (long dashes) curves, respectively. 
Cross-section values in cm$^{2}$ are given on the left y-axis.
Grey curves are Raman scattering branching ratios (right hand y-axis) with the same line-types.
The branching ratio for $n=4$ is omitted for sake of clarity.
Corresponding Raman scattered wavelengths for $n=2$ and $n=3$ are indicated on the two bottom lines.
\label{fig:fig1}
} 
\end{figure*}
This paper owes  a great debt to H.-W. Lee and S.-J.~Chang for sharing their cross-sections for Raman scattering by atomic hydrogen from its ground state to levels $n=2$, 3, and 4.
These ultraviolet cross-sections were computed for photons less energetic than $Ly\epsilon$  (938~\AA, 13.22~eV). 
For this reason the Lee/Chang database considers only optical and infrared scattered wavelengths above 4145~\AA, 1.125, and 2.809~$\mu$m for  $n=2$, 3, 4.
Above $Ly\epsilon$ and below Lyman's limit $Ly_\infty$, hydrogen Raman cross-sections (for any level) consist in a succession of close high maxima with deep minima between \citep[Figure~3 in][]{nuss89} because the energy difference between successive high Rydberg states of hydrogen tends to 0.

Raman scattering by hydrogen at rest and de-excited to level $n$ requires photons with energy greater than the transition from level $n+1$ to level $n$, $T_{n+1\rightarrow n}$. 
Raman shifts for   $n=$~2, 3, 4 are 82259, 97492, 102826~cm$^{-1}$.
The corresponding maximum wavelengths/minimum energy required for a photon to be Raman-scattered are  1215.7~\AA\ ($Ly\alpha$, 10.2~eV), 1025.7~\AA\ ($Ly\beta$, 12.1~eV),  972.5~\AA\ ($Ly\gamma$, 12.8~eV).
Photons between these wavelengths and  $Ly_\infty$ are Raman scattered in the optical and infrared spectrum from 3647, 8206~\AA\ and 1.46~$\mu$m (reached by source ultraviolet photons at the ionization limit) to infinity, for $n=2$, 3, 4.
These data are reported in Table~\ref{tbl:rcar}.

\subsection{Raman cross-sections} \label{csuv}
\begin{figure}[]
\resizebox{1.\columnwidth}{!}{\includegraphics{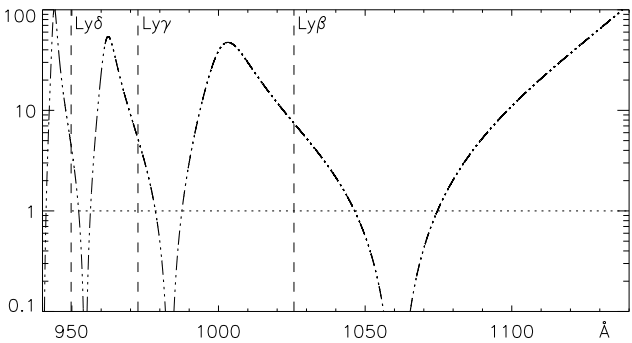}} 
\caption{Ratio of Rayleigh cross-section to the total Raman cross section (sum of cross-sections for levels $n=2$, 3, 4) in the  940--1150~\AA\ wavelength range.
Rayleigh scattering dominates over most of the spectrum.
} 
\label{fig:fig2}
\end{figure}
Ultraviolet cross-sections and branching ratios for Raman scattering leaving hydrogen at levels $n=2$, 3, 4 are plotted in Figure~\ref{fig:fig1} along with the Rayleigh and total cross-sections.
The Rayleigh cross-section dominates the Raman cross-section everywhere  except  close to Rayleigh's minima (Figures~\ref{fig:fig1} and \ref{fig:fig2}).

Raman scattering cross sections for  different  hydrogen levels (here $n=2$, 3, 4) follow a similar pattern.
The cross-section for level $n$ consists of maxima  at all Lyman wavelengths $Ly\scriptstyle{p}$ with $p>n$  and minima between.
Increasing $n$ or $p$ decreases the wavelength distance between maxima (or minima, Figure~\ref{fig:lfli}). 
From $Ly\scriptstyle{n+1}$ to $Ly\scriptstyle{n}$ the cross-section goes asymptotically to zero, with no minimum.

A too straightforward interpretation of Figure~\ref{fig:fig1} could be misleading.
Whether arising from gas or dust, large ultraviolet optical depths may reverse the apparent dominance of Rayleigh over Raman scattering (next sections).
\subsection{Gas and dust extinctions in interstellar clouds} \label{ext}
\subsubsection{Gas  extinction} \label{gext}
\begin{figure}[]
\resizebox{1.\columnwidth}{!}{\includegraphics{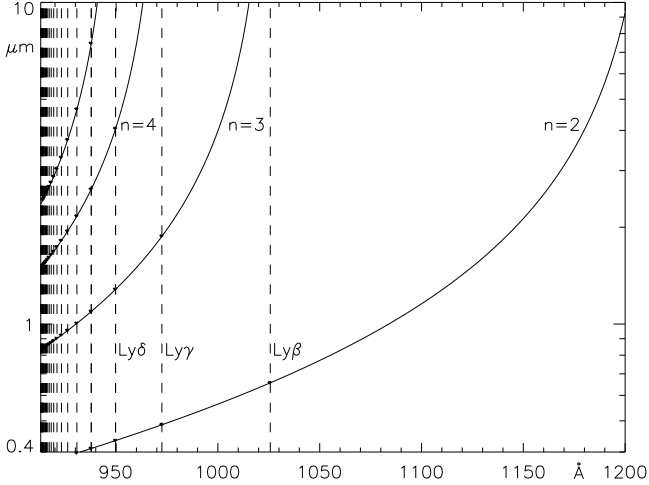}} 
\caption{Scattered $\lambda_f$ versus source $\lambda_i$ wavelengths for Raman scattering leaving hydrogen on levels $n=2$, 3, 4, 5.
The intersection of each curve with the ordinate-axis marks the minimum wavelength (maximum energy) that Raman scattering of photons less energetic than  $Ly_\infty$ can reach  for the given level $n$. 
For all $n$ the Raman scattering cross-section reaches a maximum at each Lyman wavelength, that is each time the $\lambda_f$ versus $\lambda_i$ curve crosses a Lyman line.
Cross-sections undergo one minimum between the lines.
} 
\label{fig:lfli}
\end{figure}
Negligible in the optical part of the spectrum (Section~\ref{ramray}), Rayleigh extinction  contributes significantly to the extinction of starlight  at ultraviolet wavelengths under  3000~\AA\  for HI column densities above $10^{21}$~cm$^{-2}$ \citep[][]{vogel90,vogel91,tartakinova09,skopal12}.
At wavelengths shorter than $Ly\alpha$  gas extinction is stronger  in the vicinity of Lyman transitions, where it is still  driven by Rayleigh scattering (Figures~\ref{fig:fig1} and \ref{fig:fig2}).
\citet{chang15}  show that in a gas environment with no dust, when  the Rayleigh optical depth is large and multiple scattering dominates, extinguished starlight is nevertheless all transformed into Raman scattered light. 

\subsubsection{Dust  extinction} \label{dext}
General interstellar extinction in the Galaxy at optical/near-infrared wavelengths is due to interstellar dust, which is well mixed with gas and has a $1/\lambda^p$ ($p$ of order 1) extinction law \citep{rv2, turner14}.
Dusty interstellar clouds are  optically thick to ultraviolet radiation.
In the Galaxy, dust-free clouds should only concern circumstellar gas, such as in Be stars.
Symbiotic stars and Herbig AeBe stars may or may not contain dust \citep{bopp81,allen84,meeus01, skopal17}.

If it is assumed that the optical extinction law of dust continues at ultraviolet wavelengths \citep[][and Section~\ref{con}]{uv2,an17}, the  dust ultraviolet optical thickness at wavelength $\lambda_{\mu m}$ (expressed in $\mu$m) can be  estimated by $\tau_d \simeq 1.14E(B-V)/\lambda^{1.37}_{\mu m}$ \citep[][]{turner14,an17}.
At  $1000$~\AA\ the extinction is of order  $27E(B-V)$.
A  reddening $E(B-V)\sim 0.1$~mag. --a column density $N(H)$ of order $6\,10^{20}$~cm$^{-2}$  \citep{bohlin78}-- extinguishes over 90 per cent of ultraviolet starlight.
At $E(B-V)= 0.2$~mag. over 99 per cent of ultraviolet starlight is extinguished.

Because the albedo of interstellar dust is  close to 1 \citep{henyey}, in dusty HI environments  the large ultraviolet optical depths reached at even moderate reddening $E(B-V)$ imply that ultraviolet photons are trapped in the clouds and undergo multiple scatterings until they are absorbed by dust or Raman scattered by hydrogen.
In the former case they contribute to dust heating and are released as far-infrared photons.
In the latter case they contribute to the optical and near/mid-infrared brightness of the HI medium.

Since the work of \citet{chokshi88}, it is assumed that the ultraviolet radiation in the 13.6–6 eV range (912--2000~\AA) entering HI  clouds is all transformed  into  far-infrared thermal radiation of interstellar grains in the clouds.
This integrated  power is measured in units of the "Habing field" $G_0$ ($=1.6\,10^{-6}$~W/m$^2$ or $1.2\,10^{-7}$~W/m$^2$/sr) by a coefficient that I call $\alpha$ in the rest of the paper.
The spectral distribution of the incident ultraviolet radiation field is  fixed by the color temperature of the source.
\subsection{Raman scattering approximations at low and high optical depth} \label{imin}
Hydrogen Raman scattering for hydrogen de-excited at any given level $n$ converts photons within a parent ultraviolet wavelength interval  [$\lambda_i$, $\lambda_i+\Delta\lambda_i$] into an observed optical or infrared interval  [$\lambda_f$, $\lambda_f+\Delta\lambda_f$].
Increments $\Delta\lambda_i$ and $\Delta\lambda_f$ are related by Equation~\ref{eq:br2}, $\Delta\lambda_i=(\lambda_i/\lambda_f)^2\Delta\lambda_f$.
The energies of Raman scattered and  source ultraviolet photons are in the ratio $\lambda_i/\lambda_f$.
Altogether, the power per unit wavelength ($S_\lambda$ in W/m$^2$/$\mu$m) within  [$\lambda_f$, $\lambda_f+\Delta\lambda_f$]  Raman scattered by hydrogen at rest and left at level $n$, and the source ultraviolet power per unit wavelength within [$\lambda_i$, $\lambda_i+\Delta\lambda_i$] extinguished by the Raman process, are in the ratio $(\lambda_i/\lambda_f)^3$.

It follows from Section~\ref{ext} that whether they are close to an atomic hydrogen transition or are in the presence of dust, all ultraviolet photons within  [$\lambda_i$, $\lambda_i+\Delta\lambda_i$] are converted into Raman scattered light or into far-infrared thermal emission.
Two approximations estimate the spectral irradiance (W/m$^2$/$\mu$m) of Raman scattering by hydrogen at rest and left at level $n$:
\begin {eqnarray}
S_{l}(\lambda_f)&\simeq& 
\sigma_{n} N_H\left(\frac{\lambda_i}{\lambda_f}\right)^3F_\lambda(\lambda_i)\,\,\, \scriptstyle{(\tau_g, \tau_d\,<<1)} \label{eq:low}\\
S_{m}(\lambda_f)&\simeq&\left(\frac{\lambda_i}{\lambda_f}\right)^3\frac{\gamma_n}{\Sigma_l\gamma_l}F_\lambda(\lambda_i) \,\,\,\,\,\,
\scriptstyle{(\tau_g\, or\, \tau_d\,>>1)} \label{eq:max},
\end {eqnarray}
where $F_\lambda$ (W/m$^2$/$\mu$m) is the spectral irradiance of the ultraviolet source on the cloud, and $\sigma_{n}$, $\gamma_n$ designate the cross-section and branching ratio of Raman scattering for level $n$ at wavelength $\lambda_i$.
The first approximation corresponds to a medium with no dust and a low gas optical depth.
Approximation~\ref{eq:max} assumes that all source ultraviolet photons in [$\lambda_i$, $\lambda_i+\Delta\lambda_i$] are extinguished.
Thus the approximation does not depend on hydrogen column density $N(H)$.
The formula is exact  in the immediate vicinity of Lyman wavelengths (Figure~\ref{fig:fig1}), but when the Raman cross-section is low, the formula  requires a correcting factor to account for dust absorption.

In these equations, gas and dust optical depths $\tau_g$ and $\tau_d$ are estimated at ultraviolet wavelength $\lambda_i$.
The sum $\Sigma_l$ extends to all Raman possibilities in [$\lambda_i$, $\lambda_i+\Delta\lambda_i$] and is equal to $1-\gamma_{ray}$, where $\gamma_{ray}$ is the  branching ratio for Rayleigh scattering.
Factor  $\gamma_n/\Sigma_l\gamma_l$ simplifies for $\lambda_f$ in the vicinity of $H\alpha$ because $\lambda_i$ is then  in the vicinity of $Ly\beta$, in which case $n=2$ and  $\gamma_2/(1-\gamma_{ray})=1$.
\subsection{Pattern of Raman scattered light in the high optical depth approximation} \label{patn}
\begin{figure}
\resizebox{1.\columnwidth}{!}{\includegraphics{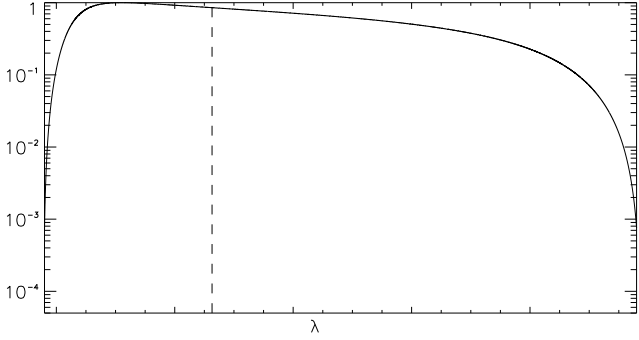}} 
\caption{Shape (from Equation~\ref{eq:max}) of Raman scattered light (scattering by hydrogen at rest and left on level $n$) in the vicinity of hydrogen transition $p\rightarrow n$ ($p>n+1$), for high ultraviolet extinction \citep[for shape variations with optical depth, see  Figure~5 in][]{chang15}.
The spectrum is normalized to 1 at its maximum.
The x-axis is the optical or infrared  interval  of the Raman scattered photons.
It is bounded  by the wavelengths that correspond to  the minima of the branching ratio $\gamma_n$ on each side of $Ly_p$.
The vertical dashed line marks hydrogen's transition $T_{p\rightarrow n}$ from level $p$  to level $n$.
For large $p$ the width is given by $0.2n^4/p^3$~$\mu$m (Equation~\ref{eq:obsdiff}).
If $p=n+1$ the right wavelength limit moves to $\infty$.} 
\label{fig:cre}
\end{figure}
 \citet[][]{chang15} used Monte Carlo simulations to investigate the profile of an ultraviolet continuum Raman scattered by hydrogen.
For wavelengths above $Ly\epsilon$, hydrogen can only de-excite to levels $n=2$, 3, or 4.
The Raman spectrum is concentrated around Balmer wavelengths  $H\alpha$ and $H\beta$ ($n=2$, Raman scattering of photons around $Ly\beta$ and  $Ly\gamma$)  and Paschen wavelength $Pa\alpha$  ($n=3$, scattering around $Ly\gamma$).
The profile of the scattered spectrum around each of these transitions resembles a hat-like shape with a width that increases with the ultraviolet optical depth \citep[Figure~5 in][]{chang15}. 
A similar profile, Figure~\ref{fig:cre}, was derived from Equation~\ref{eq:max}.

The profile  of an ultraviolet continuum  Raman scattered by hydrogen de-excited to level $n$ must resemble a series of crenels centered close to transitions $T_{p\rightarrow n}$ and separated by scattered light minima  on each side of the transitions. 
Each series is bounded on its blue side by transition $T_{\infty\rightarrow n}$ (Raman scattering of $Ly_\infty$ photons).
On the red side, a last crenel is centered close to $T_{n+1\rightarrow n}$.
The series ends by a continuum that decreases indefinitely towards the infrared.
Figure~1 in \cite{uib} marks the  $T_{p\rightarrow n}$ transitions of the  eight series ($n=4$ to $n=11$) that occupy the 3-12~$\mu$m wavelength range.
The spectrum  of the Raman scattered   ultraviolet continuum consists of all these series taken together.

Close to the blue boundary $T_{\infty\rightarrow n}$ (wavelength  $\lambda_{\infty,n}$, $\lambda_{\infty}=\lambda_{\infty,1}$) the crenels can hardly be distinguished and form a near continuum.
For large $p$ the energy and wavelength of transition $T_{p+1\rightarrow p}$ are respectively
\begin {equation}
 E_{p+1,p}\simeq \frac{31\,\mathrm{eV}}{p^3}
\label{eq:ediff}
\end {equation}
and
\begin {equation}
 \lambda_{p+1,p}\simeq \frac{2\lambda_\infty}{p^3}
\label{eq:ldiff}
\end {equation}
\begin{figure}[b]
\resizebox{1.\columnwidth}{!}{\includegraphics{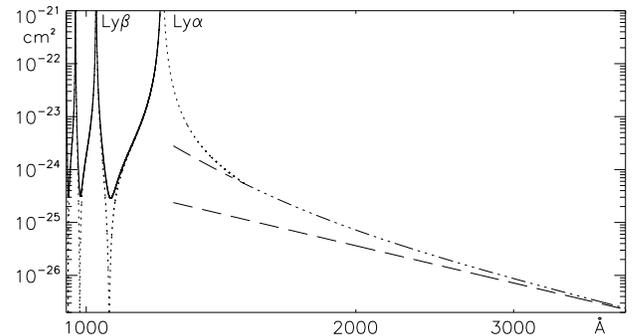}} 
\caption{Cross-section for Rayleigh scattering by hydrogen (dots), its polynomial approximations for both $\lambda>1500$~\AA\ and  $\lambda>4000$~\AA\ (the $1/\lambda^4$ law) are  in dashes.
The solid curve is the total hydrogen ultraviolet cross-section (Raman + Rayleigh, wavelengths under $Ly\alpha$).
} 
\label{fig:ray}
\end{figure}
\begin{figure*}[]
\resizebox{2.\columnwidth}{!}{\includegraphics{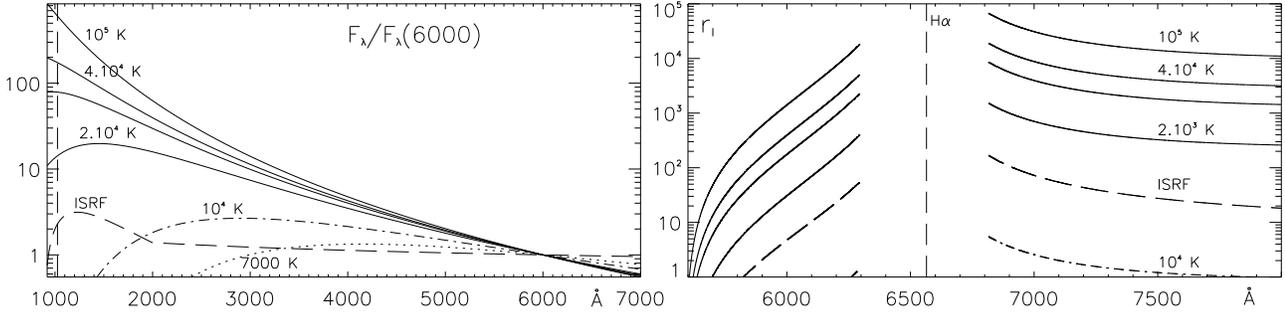}} 
\caption{Left plot: Blackbody radiation fields and mean interstellar radiation field (ISRF) in the solar neighborhood  normalized to their value at 6000~K. 
The dashed vertical line on the left marks $Ly\beta$.
The non-labeled curve corresponds to $T=3\,10^{4}$K.
Right plot: Ratio $r_l$ (Equation~\ref{eq:rlow}) around $H\alpha$ of the relative efficiencies of Raman and Rayleigh scattering in the low column density approximation, calculated  for left plot's radiation fields.
}
\label{fig:rf}
\end{figure*}
From   Equations~\ref{eq:br2} and \ref{eq:ldiff} the approximate observed crenel width  around transition   $T_{p\rightarrow n}$  is ($p>>n$)
\begin {equation}
\Delta \lambda_{obs}\simeq \frac{22}{p^3}\frac{\lambda_{\infty,n}^2}{1\mathrm{\mu m}}=\frac{0.2n^4}{p^3}\,\mathrm{\mu m}
\label{eq:obsdiff}
\end {equation}
Wavelength $\lambda_{\infty,n}$ is given by:
\begin {equation}
\lambda_{\infty,n}= \frac{n^2}{R_H}=\frac{n^2}{11}\,\mathrm{\mu m}
\label{eq:obsl}
\end {equation}
$R_H=11$~$\mu$m$^{-1}$ is Rydberg's constant for hydrogen.

\section{Contribution of Raman scattering to scattered light from HI clouds at optical and infrared wavelengths} \label{rsca}
\subsection{Raman versus Rayleigh scattering at optical  wavelengths} \label{ramray}
Practically speaking, what matters in comparing the relative efficiencies of Rayleigh and Raman scattering is not so much their relative ultraviolet cross-sections at a given ultraviolet wavelength, which generally favor Rayleigh scattering (Section~\ref{csuv} and Figure~\ref{fig:fig2}), but the number of ultraviolet photons that are Raman scattered in a given small optical/infrared interval  [$\lambda_f$, $\lambda_f+\Delta\lambda_f$] and the number of optical or infrared photons Rayleigh-scattered in the same interval.

Moving from ultraviolet to  optical wavelengths, Rayleigh scattering experiences a last maximum at $Ly\alpha$ before a continuous steep decrease (Figure~\ref{fig:ray}).
An 8$^{th}$ order polynomial reproduces the cross-section for $\lambda>1500$~\AA\ \citep[Equation~14 in][]{lee04}.
Above 4000~\AA\ the cross-section is well approximated by
\begin {eqnarray}
\sigma_{ray}&\simeq& 8.41\,10^{-25}\left(\frac{912\,\text{\AA{}}}{\lambda}\right)^4\, \mathrm{cm}^2  \label{eq:ray} \\
&\simeq& 5.82\,10^{-13}\left(\frac{1\,\text{\AA{}}}{\lambda}\right)^4\, \mathrm{cm}^2  \label{eq:ray}
\end {eqnarray}
At 6000~\AA\ the cross-section is $\sim 5\,10^{-28}$~cm$^2$.
The Rayleigh optical depth in a medium with a hydrogen column density $N(H)$ in the range $10^{19}-10^{23}$~cm$^{-2}$ is in the range $\tau_{ray}\sim 5\,10^{-9}-5\,10^{-5}$.
The Rayleigh scattered irradiance, estimated by $\tau_{ray}/(4\pi)$ times the strength of the radiation field, remains a negligible source of extinction and  brightness.

Since Raman scattering to $n=2$ (Raman cross-section $\sigma_2$) is the only hydrogen Raman scattering process that contributes to scattered starlight under 8000~\AA, ratios of approximations~\ref{eq:low}--\ref{eq:max} to Rayleigh scattered light will be (wavelengths in \AA)
\begin {eqnarray}
r_{l}(\lambda_f)&= &
1.7\,10^{12}\lambda_i^3\lambda_f \sigma_{2}  \frac{F_\lambda(\lambda_i)}{F_\lambda(\lambda_f)} \label{eq:rlow}\\
r_{m}(\lambda_f)&=& 1.7\,10^{12}\lambda_i^3\lambda_f\frac{1}{N(H)}\frac{F_\lambda\left(\lambda_i\right)}{F_\lambda\left(\lambda_f\right)}  \label{eq:rmax}
\end {eqnarray}
Radiation fields on an HI cloud can arise from a star, assumed to have a blackbody-like spectrum, or from diffuse Galactic light.
Figure~\ref{fig:rf}, left plot\footnote{The ISRF was downloaded from E. van Dishoeck's website https://home.strw.leidenuniv.nl/$\sim$ewine/photo/radiation\_fields.html. See also  \cite{heays17}.}, illustrates such radiation fields, normalized to their value at 6000~\AA.

The right side of Figure~\ref{fig:rf} plots $r_{l}$ (Equation~\ref{eq:rlow}), which is independent of column density, for these radiation fields.
The plot shows that  even in the low optical depth approximation, Raman scattered light in the vicinity of $H\alpha$ can overwhelm Rayleigh scattering by several orders of magnitude.
Since $\sigma_2$ quickly falls under $10^{-24}$~cm$^2$ on each side of $Ly\beta$ (Figure~\ref{fig:fig1}), ratio $r_{m}/r_{l}= (N(H)\sigma_2)^{-1}$ ($=\tau_{g}^{-1}$)  is much larger than 1 for all currently observed column densities.
It follows that Raman scattering must overwhelm Rayleigh scattering by several orders of magnitude for radiation field color-temperatures above $\sim10^4$~K, including diffuse Galactic light.
\subsection{Contribution of Raman scattering to the optical and infrared brightness of nebulae} \label{ramred}
Section~\ref{ramray} showed that any gas scattering  observed in the red continuum of  nebulae is Raman rather than Rayleigh scattering.
It can also be deduced that the transfer of power from the ultraviolet $Ly\beta$ to the optical $H\alpha$ regions is important enough to rival the optical illuminating radiation field at the cloud location.
The profile of the scattered light around $H\alpha$ should resemble that of  Figure~\ref{fig:cre} (with an asymptotical decrease of the red edge).  

Raman can still be expected to dominate over Rayleigh scattering under 5000~\AA, though the difference will not be as great  because of the steep increase of Rayleigh scattering (Figure~\ref{fig:ray}) and because high energy ultraviolet photons are spread in different parts of the optical/infrared spectrum  (Figure~\ref{fig:lfli}) according to the Raman branching ratios  for the different accessible levels of hydrogen.
Under 5000~\AA\ detection of Raman scattered light in nebulae is also hampered by the strong gradient of starlight (scattered in the atmosphere or on the surface of the telescope) that contaminates the observation (Appendix~\ref{sell}).
It may nevertheless be that the "blue luminescence" observed by  \citet{vijh05} in several nebulae and discussed in \citet{witt14} corresponds to near-$Ly_\infty$ photons Raman scattered close to the Balmer limit  $H_\infty$.

Towards infrared wavelengths the spectral pattern of hydrogen Raman scattered light  should be as described in Section~\ref{patn}, with narrower series of crenels as hydrogen's final state $n$ increases. 
\section{Scattering from HI media and geometry} \label{scage}
\subsection{One and two-step processes in Raman scattering} \label{sym}
In all  Raman emission line detections reported to date, the source  ultraviolet lines are  created prior to being Raman scattered by HI gas, usually in the HII region located between the source radiation field and the HI region.
These lines also leave absorptions in the continuum of the source radiation field and broad absorptions in the continuum's Raman scattered spectrum \citep{henney21}.
A complete observed Raman spectrum must thus comprise  Raman emission lines and the Raman scattered continuum of the source radiation field with absorption dips at the position of the emission lines.

This mingling of Raman scattered direct (continuum + absorptions) and re-processed (emission lines) light could provide an explanation of the double peak structure \citep[two peaks separated by a dip, see Figures~1 in ][]{allen80,schmid89} often observed in Raman emission lines.
Up to now this structure has  been justified either by self-absorption of the original emission line before Raman scattering \citep{schwank97}, or by a velocity difference  \citep[Doppler effect,][]{schmid96,heo15} in the HI and HII media involved in Raman scattering \citep[see also][]{lee00b}.
But the double peak structure could also result from the superposition of  Raman scattering of the emission line on  Raman scattering of  the source continuum with the corresponding line in absorption. 
\subsection{Dust and gas scatterings: analyzing scattered starlight with respect to scattering angles } \label{sca}
Gas (over 90 per cent hydrogen) and interstellar dust are the only confirmed components of HI clouds.
Interstellar dust is negligible in quantity but known for its strong extinction capacity and for its strong forward scattering phase function \citep[][and Section~\ref{ext}]{henyey}.
Interstellar dust scattering is observed when the angle of scattering is small and thus when the source of illumination is far behind the scattering medium.
This is the case with high latitude cirrus illuminated by a bright star,  by general background starlight, or by a galaxy \citep{pol,neb,juvela08}.

If gas scattering is observed in light reflected by interstellar matter at optical and infrared wavelengths it will be  Raman scattering and will be observed at localized wavelength ranges centered on the Balmer lines in the optical spectrum and on the long wavelengths of hydrogen transitions in the infrared (Sections~\ref{patn}, \ref{rsca}, \ref{uibl}).
In contrast to dust scattering, gas scattering is nearly isotropic.
At optical wavelengths, the loss of power that results from this isotropy is compensated by the proximity of the source of light (e.g. one or several stars).

The general rule that dust and gas scattering can be differentiated by the angle of scattering suffers one exception: scattering in the near-complete forward direction by identical particles.
This occurs when scattered and direct starlight are combined in the same beam (the star is observed through the scattering HI medium).
In this case H atoms  no longer behave as separate particles but  create a coherent wave that amplifies the scattering.
At any given wavelength, the sustainability of this wave depends on the size of the first Fresnel zone at the cloud location as viewed by the observer, and thus on the distances between the observed star, the HI scattering cloud, and the observer.
The scattering irradiance varies as  $N(H)^2$ and can be considerably  greater than forward scattering by dust grains.
Coherent Raman scattered light can be stimulated by the star's own optical radiation field, which is also coherent at the optical Raman-scattered wavelengths.
In stimulated Raman scattering (SRS) the out-going Raman scattered light depends on the ultraviolet and on the optical energy distributions of the radiation field \citep{prince16}.
\subsection{Polarization of extinguished and scattered starlight} \label{pol}
Scattered light is generally polarized.
The polarization increases with the angle of scattering (no polarization in the frontward direction because of the symmetry), and decreases with increasing wavelength for any given particle.
For particles of size $a$  polarization varies as $a/\lambda$.

Observation shows that  in the optical spectrum, light from a star extinguished by interstellar matter does not steadily decrease with increasing wavelength but follows a Serkowski law \citep{serkowski75}.
It  increases with wavelength in the blue and decreases in the red, with a maximum between 4500 and 8000~\AA.
This shape must result from the size distribution and shape of interstellar dust. 

It follows that the dependence of the polarization on wavelength differs for each of the three different types of scattering discussed in Section~\ref{sca}. Scattering by dust  follows a Serkowski law, growing with wavelength in the blue and decreasing in the near-infrared part of the spectrum.
If scattering by gas is observed,  polarization must increase  with decreasing wavelength.
In the case of complete forward scattering, scattered starlight  is not polarized.
\section{Extended Red Emission (ERE)} \label{ere}
\begin{table*}
\begin{center}
\caption{ERE observed and expected measures. 
For each object  columns 5--6 give the calculated (see Section~\ref{ere} and the  Table's notes) and  observed ERE brightness.
Last two columns compare the ultraviolet power close to $H\alpha$ needed to produce the observed ERE in-band power (column~4) and the source's ultraviolet power within 1010 -- 1045~\AA\  deduced from far-infrared thermal emission ($\alpha G_0$) and color temperature of the source radiation field.
Observed values are in bold.
\label{tbl:ere}
}		
\begin{tabular}{l|rrr|rr||rrr|}
object& $T$& $\alpha^{(1)}$ & $\overline{F_{Ly\beta}}\,\,^{(2)}$ & $I_{ere}^{calc}\,^{(3)}$ &{ \boldmath{$I_{ere}^{obs}$}} $\,^{(4)}$ & {\boldmath{$P_{ere}$}} $\,^{(5)}$ & $P_{Ly\beta}^{Ram}$ $\,^{(6)}$ & $P_{Ly\beta}^{dust}$ $\,^{(7)}$\\
 & \scriptsize{K} &  &\scriptsize{W/m$^2$/$\mu$m}& \scriptsize{W/m$^2$/$\mu$m/sr}& 
 \scriptsize{\boldmath{W/m$^2$/$\mu$m/sr}} & \scriptsize{\boldmath{W/m$^2$/sr}} & \scriptsize{W/m$^2$/sr}  & 
 \scriptsize{W/m$^2$/sr } \\
\hline
NGC7027$^{(a)}$ & $2\,10^5$&   $2\,10^5$ &10\,\,\,\,\,\,& $3\,10^{-3}$ &\boldmath{$7\,10^{-4}$} & \boldmath{$10^{-4}$} & $6.4\,10^{-4}$ &  $2.2\,10^{-3}$\\ 
\hline
NGC7023$^{(b)}$ & $2\,10^{4}$ & $10^{3}$ &$1.6\,10^{-2}$ & $4.3\,10^{-6}$ &\boldmath{$2\,10^{-6}$} & \boldmath{$2\,10^{-7}$}  &$1.3\,10^{-6}$ & $4.7\,10^{-6}$\\ 
 \hline
 NGC2023$^{(c)}$& $2\,10^{4}$ &  $2\,10^{2}$& $2.9\,10^{-3}$ & $8.6\,10^{-7}$  &\boldmath{$10^{-6}$} & \boldmath{$1.2\,10^{-7}$} &$7.7\,10^{-7}$  & $9.4\,10^{-7}$\\ 
 \hline
RR 6"$^{\,\,(d)}$&  $10^{4}$& $1.4\,10^{4}$ &$2\,10^{-2}$& $7\,10^{-6}$$^{\,\,(d)}$ &\boldmath{$4\,10^{-5}$} & \boldmath{$7\,10^{-6}$} & $4.5\,10^{-5}$ &  $4.6\,10^{-6}$$^{\,\,(d)}$\\ 
RR 10"$\,\,^{(d)}$& $10^{4}$&$5.6\,10^{3}$ &$6\,10^{-3}$&$2\,10^{-6}$$^{\,\,(d)}$ &\boldmath{$1.5\,10^{-5}$}& \boldmath{$10^{-6}$} & $6.4\,10^{-6}$ & $1.8\,10^{-6}$$^{\,\,(d)}$ \\ 
\end{tabular}
\end{center} 
\begin{list}{}{}
\item[$(1)$] Estimated ultraviolet irradiance on the cloud in units of $G_0\,=\,1.6\,10^{-6}$~W/m$^2$.
\item[$(2)$]Spectral irradiance at $Ly\beta$ at the cloud position (see text, Section~\ref{ere}). 
\item[$(3)$] ERE calculated from $\overline{F_{Ly\beta}}$ and Equation~\ref{eq:ere0} (NGC7027, Red Rectangle) or Equation~\ref{eq:ere1} (NGC7023, NGC2023).
\item[$(4)$] Approximate observed peak ERE spectral radiance. All values except for NGC7027 are from \citet[][Table~2]{witt90}.
\item[$(5)$] Observed in-band ERE power. All values except for NGC7027 are from \citet[][Table~2]{witt90}.
\item[$(6)$] In-band ultraviolet source power around $Ly\beta$ needed for Raman scattering to cause ERE: $P_{Ly\beta}^{Ram}=(\lambda_f/\lambda_i)${\boldmath${P_{ere}}$} $=6.4${\boldmath${P_{ere}}$}.
\item[$(7)$] Fraction of $\alpha G_0$ emitted  in the 1010 -- 1045~\AA\ wavelength range (Raman scattered between 6000 and 7500~\AA), assuming that  all ultraviolet wavelengths contribute equally to far-infrared emission.
\item[$(a)$] Temperature from  \cite{zhang05}, {\boldmath{$I_{ere}^{obs}$}} and  {\boldmath{$P_{ere}$} } estimated from \citet[][Figure~1]{furton90}
\item[$(b)$] $\alpha$ from \citet[][$\alpha=2.6\,10^3G_0$ 35" from HD200775]{chokshi88}, rescaled to 54", the distance of Witt \& Boroson's observation.
\item[$(c)$] $\alpha$ from \citet[][$\alpha=3\,10^2G_0$ 1 arc-minute from HD37903]{hony01}, rescaled to 70", the distance of Witt \& Boroson's observation.
\item[$(d)$] Data are derived from estimated temperature, luminosity, and distance of HD44179, $T=10^4$~K,  $L_\star=10^3L_\odot = 4\,10^{29}$~W, $d=330$~pc \citep{geballe89}. Note that $\alpha$ values of column~2 are low compared to the value derived from the far-infrared brightness of HD44179  by \citet[][$\alpha\,=\,5\,10^6$]{hony01}, which would yield $\alpha$-values on the order of 17 times higher (S.~Hony, private communication). 
\end{list}
\end{table*}
ERE is known to  require a radiation field with a color temperature higher than about $10^4$~K  \citep[no ERE is observed when the illuminating star has a temperature less than 7000~K,][]{darbon99,witt20}.
\citet{smith02} further found a close correlation between the intensities of ERE and of the far-ultraviolet radiation field.
The  low resolution ERE profile of NGC7027 in \citet[][their Figure~2]{furton90} strikingly resembles the hydrogen Raman scattering Monte Carlo simulations of \citet[][top plot of their Figure~3]{chang15}.
In general, the blue ERE edge rises  shortly before 5600~\AA\ \citep{witt90,perrin92,vanwinckel02}, close to the reddest minimum of the cross-section of hydrogen Raman scattering (Figure~\ref{fig:fig1}, this minimum is  at  5580~\AA, 998~\AA\  in the parent ultraviolet spectrum).
ERE peaks in the $H\alpha$ region, between 6000 and 7000~\AA\ \citep{witt14}.

From Equation~\ref{eq:max}, the maximum spectral radiance (W/m$^2$/$\mu$m/sr) ERE can reach is
\begin {equation}
I_{ere}=\frac{1}{4\pi}\left(\frac{\lambda_i}{\lambda_f}\right)^3\overline{F_{\lambda_i}}=3\,10^{-4}\overline{F_{Ly\beta}} \label{eq:ere0}
\end {equation}
where $\overline{F_{Ly\beta}}$ (W/m$^2$/$\mu$m) is the mean value of the source radiation field around $Ly\beta$ at the cloud location, and $\lambda_i$ and $\lambda_f$ are the $Ly\beta$ and $H\alpha$ wavelengths ($\lambda_i/\lambda_f=0.156$).

For illuminating radiation fields such as NGC7023 or NGC2023 with a color temperature close to that of van~Dishoek's interstellar radiation field \citep[$T\sim 2\,10^4$~K,  Figure~\ref{fig:rf} and][]{heays17}, Equation~\ref{eq:ere0} can be re-written
\begin {equation}
I_{ere}\simeq 4.3\,10^{-9}\alpha  \,\,\,\,\,\,\,\,\,\,\,\,\,\,\,\,\,\, \,\,\,\,\,\,\,\,\,\,   \left(\alpha=\frac{\int_{UV}{F_\lambda}d\lambda}{G_0}\right) \label{eq:ere1}
\end {equation}
$G_0$ and $\alpha$ were defined in Section~\ref{dext}.
Van~Dishoeck's ISRF integrated from 912 to 2000~\AA\ is equal to 1.6$G_0$ and its  value  at $Ly\beta$ is $2.3\,10^{-5}$~W/m$^2$/$\mu$m.

In NGC7027 the ERE plateau observed by Furton \& Witt stands at  $7\,10^{-4}$~W/m$^2$/$\mu$m/sr \citep[Figure~2 in][]{furton90}.
The ultraviolet radiation field  on the HI scattering medium  is at least $2\,10^5G_0$ \citep[estimated by][]{hony01}.
For a star temperature of the order of $2\,10^5$~K \citep{zhang05}, the radiation field at $Ly\beta$ on the cloud should be close to 10~W/m$^2$/$\mu$m.
From Equation~\ref{eq:ere0},  the estimated plateau of ERE/Raman scattered light is at $3\,10^{-3}$~W/m$^2$/$\mu$m/sr, the same order of magnitude as observed by Furton \& Witt.

Columns~5 and 6 of Table~\ref{tbl:ere} compare the ERE peak expected from Equation~\ref{eq:ere0} (Equation~\ref{eq:ere1} for NGC7023 and NGC2023) to the observed peaks for NGC7023, NGC2023, and two positions in the Red Rectangle.
The Table shows that  Raman scattering can account for the observed ERE peaks for the estimated radiation fields at the cloud locations.

A comparison of ERE spectra in \citet{witt90} and \citet{furton90} with Figure~5 in  \citet[][]{chang15} indicates that the widths of observed ERE are  too large to result from gas extinction alone (too high hydrogen column densities would be required).
This  conclusion supports the possibility advanced in Section~\ref{dext} that dust extinction contributes to keeping ultraviolet photons locked in  HI media and thus complements hydrogen in transferring energy from ultraviolet to optical wavelengths.
On the assumption that dust absorption is equally distributed among ultraviolet wavelengths, the last two columns of Table~\ref{tbl:ere}  compare  the source power (W/m$^2$/sr) around   $Ly\beta$ needed to produce ERE (roughly the integrated source power  between 1010 and 1045~\AA), deduced from the observed ERE integrated in-band power, to the power absorbed by dust between 1010 and 1045~\AA. 
This comparison could indicate that Raman scattering and dust absorption contribute in a similar proportion  to the transfer of energy from  the vicinity of $Ly\beta$ to the vicinity of $H\alpha$. 

In the rest of the paper I take it for granted that ERE is indeed Raman scattering by hydrogen.
This account of ERE provides a more natural explanation  than the complex two-step PAH emission process imagined by \cite{witt06}, the "recurrent Poincar\'e fluorescence" of \cite{leger88,perrin92, witt20}, or any other such emission process that would require 300 per cent efficiency \citep[][]{gordon98}.
\section{From Extended Red Emission to Diffuse Interstellar Bands} \label{ed}
The Red Rectangle nebula  has a spectacular emission-like spectrum that takes the form of several series of relatively broad crests and valleys on top of the ERE  \citep[Figure~1 in][]{schmidt80}.
These abnormally broad features are distributed on four continuum enhancements that occupy roughly the 5800-5950, 6150-6300, 6300-6480, 6550-6700~\AA\  regions \citep[Figures~2 and 3 in][]{vanwinckel02}.
A dozen among some 50 reported bands are particularly strong, giving the Red Rectangle spectrum its specific character \citep[Table~4  in][]{vanwinckel02}.

 \citet[][see their Table~5]{vanwinckel02} found that nine of the strong Red Rectangle bands were situated on the red edge of  the strongest DIBs (such as DIBs~5797, 5850, 6196, 6203, 6614).
Re-examination of Van~Winckel et al.'s   work in light of  recent DIB databases  shows that almost any of their  Table~4's four dozen Red Rectangle bands  can potentially match a DIB.
For instance, the relatively strong Red Rectangle emission band at 6378.6~\AA\ in the 6300-6480 complex \citep[Figure~3 in][]{vanwinckel02}  is flanked by DIB~6377 on its blue side.

Despite the undeniable correspondence between DIBs and  Red Rectangle emission bands, the possibility that they arise from a common carrier has been questioned on the ground that central   wavelengths of the Red Rectangle bands remain 1 to 2~\AA\ redshifted from their associated DIB central wavelength \citep{glinski02,vanwinckel02}.
However, if we assume that ERE is starlight Raman scattered by atomic   hydrogen, this 1-2~\AA\ shift reduces (by a factor $6.4^2$) to 0.02-0.05~\AA\ in the parent ultraviolet spectrum.
In fact, both \citet[][Figure~1]{witt14}, who remarked that DIBs and ERE have a similar spectral density distribution, and  \citet{lai20}, who observed DIBs and ERE in nebula IC63 (Section~\ref{ic63}), concluded that DIBs and ERE are related.
I also note that all but two DIBs ($\lambda\lambda$6645.53 and 6709.24) of van~Winckel et al's Table~5 can be associated on their blue side with a small Red Rectangle band of their Table~4.

As anticipated in Section~\ref{sym}, these connections between ERE and DIBs can be explained if the ERE spectrum consists in the mingling of Raman scattering  of the continuum close to $Ly\beta$ with Raman emission lines.
A strong implication of this conclusion is  that the spectrum of a star with observed DIBs also includes a  component of Raman scattered starlight.
Because it is observed in the direction of the star, the scattering must be coherent (Section~\ref{sca}).
A component of scattered light in the spectrum of reddened stars also justifies the two-parameter dependency of  ultraviolet extinction curves \citep{an17,param}.
\section{Diffuse Interstellar Bands} \label{dib}
\begin{figure}[b]
\resizebox{1\columnwidth}{!}{\includegraphics{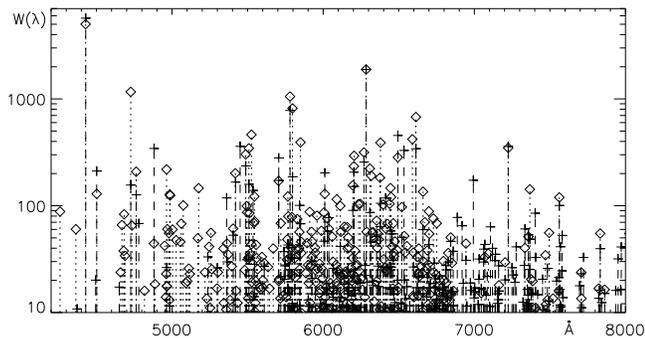}} 
\caption{DIB spectrum (equivalent widths, EWs) of HD183143 (+) and HD204827 (diamonds, EWs are multiplied by 7).
From \citet[][]{hobbs08} for HD204827 and \citet{hobbs09} for HD183143.
DIBs may not be in the same exact proportion in both stars,  but the DIB spectrum  has a defined pattern with a large concentration of DIBs between 5700 and 6800~\AA, and roughly  centered on DIB~$\lambda6284$ \citep[see also Figure~1 in][]{witt14}.
Raman scattering cross-section for level $n=2$ of hydrogen has two minima under $Ly\epsilon$ close to $960.6$ (observed at  5580~\AA) and 998~\AA\  (observed at  5780~\AA). 
$\lambda 5580$ is clearly a region devoid of DIBs.
Enlargement of the x-axis would also reveal a minimum of DIB concentration at  $\lambda 5780$.
} 
\label{fig:dib}
\end{figure}
\subsection{Properties of the Diffuse Interstellar Bands and Raman scattering} \label{dibor}
DIBs are generally broad \citep[Figure~7 in][]{hobbs09}, a characteristic  of Raman scattering (Sections~\ref{intro}--\ref{rs}).
The area of the first Fresnel zone, and thus the number of hydrogen atoms that contribute to coherent Raman scattering, diminishes as the distance between the star and the obscuring material  decreases \citep[][and Section~\ref{sca}]{an17}.
This dependence on distance of the size the first Fresnel zone accounts for the tendency for DIBs  to vanish when the interstellar matter is in the vicinity of stars.
The branching ratio for level $n=2$ has two minima at  $\lambda\simeq 960.6$ and 
998~\AA\  (Figure~\ref{fig:fig1}), which in the space of observation correspond to $\lambda\simeq 4578$ and 5580~\AA, two wavelength regions with a low DIB concentration  (Figure~\ref{fig:dib}).

DIBs can be divided  into strong and weak DIBs.
Strong DIBs, such as $\lambda\lambda$~4440, 5780, 5797, 6196, 6284, 6613, clearly stand out on a DIB spectrum (Figure~\ref{fig:dib}).
These DIBs seem to extinguish most, if not all, of the scattered component of the stars' spectrum.
I considered the possibility that they could be related  to atomic hydrogen itself or to HeII, which predominate in HII and photodissociation  regions, but I have not undertaken a deeper investigation.

The observed separation between the wavelengths of two DIBs shrinks in the source ultraviolet spectrum.
In the parent space DIBs~5780 and 5797 correspond to wavelengths 1004.4~\AA\ and 1004.9~\AA, which are only 0.5~\AA\ apart.
Broad DIB~4428, which occupies the 4400--4450~\AA\ wavelength range, is the Raman scattered counterpart of the 1~\AA\ wide 953.8--954.8~\AA\  interval.

The vast majority of DIBs are weak DIBs.
Their number has  increased steadily with the increasing sensitivity of observations.
The original ultraviolet absorptions must occur in the stars'  light as it enters an interstellar cloud, either in  its HII outskirts (Schmidt's OVI lines, for instance)   or inside the PDR  (Henney's OI lines), before the starlight is Raman scattered.
\subsection{Identification of  Diffuse Interstellar Bands} \label{dibid}
In the most recent \cite{fan19} catalog, DIBs~6827 and 7084 (n$^o$~426 and 463 in the catalog)  fall at the exact wavelengths of the Raman scattered  OVI lines observed by \citet{schmid89}.
Their equivalent widths  \citep[29.7 and 6.3~m\AA\ in][]{hobbs09} are also in the expected 4:1 ratio given by Schmid.

The OI ultraviolet doublet  is Raman scattered at 6633 and 6664~\AA, DIBs n$^o$~358 and 370 in Fan et al.'s catalog\footnote{ \cite{henney21} noted the strongest of the OI absorptions in a DIB catalog, but he missed the weakest one (not listed in the catalog) and denied a correlation.}.
The DIBs' equivalent widths are in the ratio 21.6:7.6 consistent with the observations of \cite{henney21}.

A difficulty in matching ultraviolet absorptions with DIBs comes from Equation~\ref{eq:br2}, which transforms a small uncertainty on the source ultraviolet wavelength into a very large one on the observed DIB spectrum.
This obstacle can be overcome by additional constraints such as those imposed by doublets.
However, I note that another DIB, $\lambda$6543.29, falls at the exact position of the strongest ultraviolet Raman scattered lines of HeII  \citep[close to 1025.24~\AA, Table~2 in][]{lee06}.
Other potential identifications, with HeII, CII, or SiII for instance, require, as for the strong DIBs, more thorough investigations.
\section{ Unidentified infrared bands (UIBs) and infrared continuum excesses} \label{uib}
\subsection{Infrared excesses, UIBs, and interstellar dust} \label{uibext}
Infrared excesses are common in space; they are also found in directions without UIBs.
UIBs are not observed in Be stars, for instance, and infrared excesses in these stars   can be explained by free-free emission \citep{woolf70,gehrz74,hackwell74,briot77,berrilli92}, even  for low star temperatures  \citep[down to 3000~K,][]{dyck72,milkey73}.

The  correlation of UIBs with dust has been systematically noticed \citep[for instance][for clouds in the Galactic plane]{kahanpaa03}.
When there is no dust (such as in Be stars) UIBs disappear.
UIBs have been detected in  about half of Herbig AeBe stars,  precisely those with a component of warm interstellar dust  \citep{meeus01,alecian11}.
In symbiotic stars UIBs have been observed  only in (dusty) D-symbiotics \citep{schild99,angeloni07}.
The presence of interstellar dust thus seems to be an essential condition for observing UIBs.

In dusty environments the term  "infrared excess" encompasses UIBs and their underlying continuum.
Their relative proportion in different directions is not as stable as the the relative proportions of UIBs alone (UIB~3.3 excepted, Sections~\ref{uibl} and \ref{uibc}).
The separation of UIBs and continuum excess in a spectrum  is not always easy to achieve \citep[for instance][]{peeters17}.
The ability to do so  depends  in part on the sensitivity of observations:
weak additional  bands can be resolved in the continuum of the 3.3--3.6~$\mu$m wavelength range \citep{grasdalen76,geballe85,demuizon86,nagata88,beintema96}, which have been shown to correlate with  the strong UIBs at longer wavelengths \citep[][]{lu03}.
These bands are thus more related to UIBs than to the continuum.
\subsection{Ultraviolet illumination in UIB observations} \label{uibil}
Just as for  ERE, in the vast majority of observations UIBs are associated with ultraviolet-dominated radiation fields with color temperatures greater than $\sim10^4$~K.
This UIB-dependence on ultraviolet light has been contested based on the detection of UIBs in vdB133, a nebula immersed  in the ultraviolet-deficient radiation field of stars HD195593A/B \citep[ultraviolet luminosity under 2000~\AA, less than 1 per cent of total luminosity, see][]{uchida98,uchida00}.
The case of vdB133  led \cite{li02}  to consider the possibility that UIBs could be formed from  optical illumination alone.
However,  vdB133 lies in the Galactic plane ($l=76.4$, $b=-1.45$), and its 7.7~$\mu$m emission, of the order of $10^{-6}$~W/m$^2$/$\mu$m/sr, has the typical value of HI clouds close to the plane and  illuminated by ambient Galactic light \citep{kahanpaa03}.
Furthermore, \citet{uchida00} are unable to detect UIBs in  nebulae vdB101, vdB11, and vdB135, all   located out of the Galactic plane ($|b|>20$) and having a low ratio of ultraviolet to optical illumination but one still larger than vdB133's.
The  vdB133 exception thus cannot support the claim that UIBs do not require far-ultraviolet photons.

UIB-like spectra of Class~C \citep[in the classification of][]{peeters02} have nonetheless been found in three S-type AGB stars with temperatures as low as 3300~K  \citep{smolders10,smolders12}. 
This type of spectrum, however, is uncommon and differs significantly from the usual UIB spectra.

\subsection{The spatial overlap of optical (ERE) and infrared excess brightness in nebulae} \label{uibneb}
As mentioned in the introduction ERE, DIBs, and UIBs, also infrared continuum excesses, peak in HI photodissociation regions.
Observing that  the spatial distribution of the optical and near-infrared brightness of NGC7023, NGC2023, and NGC2068 correspond, \citet{sellgren83t,sellgren84}  suggested that the optical and near-infrared brightness of nebulae had  a common carrier.
Under the assumption  that the optical brightness of nebulae resulted from Rayleigh scattering of starlight by small grains, she further wondered  whether the near-infrared excess could not be the tail of the optical scattering.
Her negative conclusion led to the well-known "transiently heated small grains" hypothesis, unfortunately forged on an inappropriate model of interstellar dust and on misinterpretations of polarization data (Appendix~\ref{sell}).

\citet[][]{tuthill02}  observed the  Red Rectangle nebula with unprecedented high resolution (46--68~mas) images in different wavelength bands and similarly concluded that
\begin{quote}
"[a] remarkable feature of the Red Rectangle bipolar nebula is its self-similar appearance on scales from 50~mas to 1' and from the red (visible) to at least 10~$\mu$m."  \citep[][p.~895]{tuthill02}.
\end{quote}

The association of ERE, UIBs, and infrared excesses with HI media and  the observations of \citet{sellgren84} and of  \citet{tuthill02} indicate that these optical/infrared features arise (within the constraints of current spatial resolution of observations) from the same HI region.
\cite{witt14} contested this spatial self-similarity of  nebulae at different wavelengths, although he omitted to cite the  \citet{sellgren84} and   \citet{tuthill02} papers and acknowledged that ERE, like UIBs and continuum excess, arise from HI photodissociation regions \citep{witt20,lai20}.
He backed up his claim with a few papers \citep[including  an AAS meeting abstract,][unsupported by a relevant publication]{kurth13} none of which  uses observations that rival those of Tuthill et al. in spatial resolution.
Rather than undermining Sellgren and Tuthill et al.'s conclusion that ERE and infrared excesses arise from the same regions, these papers simply indicate that ERE and infrared excesses may not remain in the same proportion from one direction to another \citep[compare, for instance, the ERE and mid-infrared maps of Figures~8 and 9 in][]{darbon00}.
\subsection{UIBs and hydrogen emission lines} \label{uibl}
\cite{merrill75} mentioned that UIB~3.3 is close to $Pf\delta$ and stands at the exact position of  $T_{\infty\rightarrow 6}$ but they dismissed a correlation on the ground that $Hu_{\infty}$ would be "significantly larger than predicted by recombination theory".
UIB~3.3  is strong in high color temperature ultraviolet radiation fields, competing in strength with UIB~11.3 \citep[Figure~1 in][]{peeters04} and is hardly detected in the Galactic plane or in galaxies \citep{kahanpaa03,lu03}. 
When observed, the band is systematically associated with several lines of hydrogen and disappears when  no hydrogen line is observed \citep{mori12}.
A spectral overlap between the weak 3.3--3.6~$\mu$m bands and the series of hydrogen Humphrey lines was further remarked and  discussed in some detail by \citet[][p.~503 on observations of NGC7027]{geballe85}, but for the same reason as Merrill et al. they also  denied a correlation.

However, in two out of the three directions, NGC7027 and the Red Rectangle,  of Geballe et al.'s study, ERE is observed and spatially coincides with the near-infrared emission (Section~\ref{uibneb}).
Also, with the exception of UIB~8.6, footprints on the spectrum of all main infrared bands and the weaker 3.3--3.6~$\mu$m  bands overlap the near continua of hydrogen transitions from high Rydberg states to lower levels $n=6$, 8, 9 and 11 \citep{uib}.
Hydrogen transitions  to $n=7$ fall within a series of weak bands between 4.5 and 5~$\mu$m \citep{verstraete96,tokunaga97,lu03}.
The weak 9-10~$\mu$m  emission observed by \citet[][ in Ophiuchus cloud LDN 1688]{rawlings13} may correspond to the hydrogen transitions to $n=10$.

These findings highlight a relationship between the spectrum of hydrogen and the UIB spectrum that  extends over the whole near to mid-infrared spectrum and goes well beyond a single coincidence in the 3.3~$\mu$m region.
It is remarkable that in nebulae where ERE is observed the imprint of the optical/infrared spectrum follows the spectrum of hydrogen, and that in both shape and strength the ERE feature agrees with expectations from Raman scattering near $H\alpha$ (Section~\ref{ere}).
I thus cannot follow \citet{merrill75} and \citet{geballe85}, and  consider  that UIB~3.3 is indeed $T_{\infty\rightarrow 6}$ and that UIBs~6.3, 7.7, and 11.2 along with the 3.3--3.6~$\mu$m sub-bands  correspond to de-excitations of hydrogen  near the ionization limit.
The spatial coincidence of ERE and infrared excesses in nebulae (Section~\ref{uibneb}) and the tendency for UIBs to huddle around hydrogen infrared transitions justify  considering the UIB spectrum as the continuation of the ERE/Raman spectrum at infrared wavelengths. 
The hypothesis that UIBs are Raman scattered starlight by hydrogen  is further supported by the involvement of dust extinction in UIB observations (Section~\ref{uibext}).
\subsection{UIBs and Raman scattering} \label{uibr}
The main UIBs~6.2, 7.7, 11.3 and the weaker 3.3--3.6~$\mu$m bands must  result from de-excitation of hydrogen atoms  on, or close  to, high Rydberg states\footnote{\cite{holmlid00}  suggested that UIBs carriers could be Rydberg matter (atoms or molecules excited to high Rydberg states) and mentioned atomic hydrogen among the possible carriers.}, down to levels $n=6,$ 8, 9, 11.
For excitation of hydrogen at rest up to a virtual state just under the ionization limit, a UIB width $\Delta\lambda_f$ ($\lambda_f$ is the UIB central wavelength)  and its counterpart   $\Delta\lambda_i$ in the parent ultraviolet spectrum are related by (Equation~\ref{eq:br2})
\begin {equation}
\Delta\lambda_i\simeq
\left(\frac{912\,\text{\AA{}}}{\lambda_f}\right)^2\Delta\lambda_f
\label{eq:duv}
\end {equation}
Adopting\footnote{These values are UIBs' approximate footprint on the spectrum \citep[][]{uib}. For the 3.3--3.6~$\mu$m bands see \citet{nagata88}. They are not to be confused with FWHM which are of order 0.24, 0.76, 0.27~$\mu$m for  UIBs~6.2, 7.7, 11.3 in the diffuse Galactic interstellar medium \citep{kahanpaa03,sakon04}.} $\Delta\lambda_f\simeq$~0.4, 0.6, 0.8, and 0.8~$\mu$m in this Equation for the 3.3-3.6~$\mu$m region and UIBs~6.2, 7.7, and 11.3, one finds $\Delta\lambda_i$-values equal to 2.8, 1.3, 1.1, and 0.5~\AA.
These values imply that hydrogen at rest must be excited to levels above $n=$~20, 27, 30, and 45.

\citet{nagata88} studied in detail the sub 3.3--3.6~$\mu$m bands in PNe NGC7027 and IRAS~21282+5050 ($\alpha\,=\,10^5$, $T\,=\,3.3\,10^4$) from ground-based spectral observations with 2.7" spatial resolution.
For NGC7027 they found for the set of sub-bands a total in-band power of $10^{-5}$~W/m$^2$/sr.
If caused by Raman scattering of ultraviolet photons by atomic hydrogen at rest, this value requires a source power of $10^{-5}\times \,(3.5\,\mu\text{m}/Ly_{\infty})\,=\,3.6\,10^{-4}$~W/m$^2$/sr,  produced by photons within 2.8~\AA\ under $Ly_\infty$.
The spectral radiance at 912~\AA\ on NGC7027's HI region is of the order of 10~W/m$^2$/$\mu$m at $Ly\beta$ and should be of the order of 15~W/m$^2$/$\mu$m ($1.5\,10^{-3}$~W/m$^2$/\AA) close to $Ly_{\infty}$.
The available source power within 2.8~\AA\ of $Ly_{\infty}$ is therefore $4.2\,10^{-3}$~W/m$^2$ or $3.3\,10^{-4}$~W/m$^2$/sr, in conformity with expectations.
A similar agreement between observation and expectation is found for IRAS~21282+5050.

The in-band power of  main UIBs~6.2, 7.7, and 11.3 is, however, much larger than in the 3.3--3.6~$\mu$m bands and cannot be accounted for by Raman scattering by hydrogen at rest of photons just under the Lyman limit.
In NGC7027 and for UIB~7.7, \citet{beintema96} found an in-band power of $2.6\,10^{-4}$~W/m$^2$/sr, which would  require a  source radiation field of $2.2\,10^{-2}$~W/m$^2$/\AA/sr at 912~\AA.
This is two orders of magnitude greater than the radiation field calculated  in the previous paragraph ($1.5\,10^{-3}$~W/m$^2$/\AA\  $\equiv 10^{-4}$~W/m$^2$/\AA/sr).

The process that gives rise to main UIBs~6.2, 7.7, and 11.3 must thus be envisaged differently from Raman scattering of ultraviolet photons just under  the Lyman limit.
Two possibilities may account for these UIBs. 
One is excitation of hydrogen by Lyman continuum (LyC) photons (created in the ionized gas at the edge of the photodissociation region) followed by recombination in an excited state close to $Ly_\infty$ and de-excitation to levels $n=8$, 9, and 11.
The other is Raman scattering by hydrogen already excited to $n=2$ or above.
In this case the atom can be excited (by an optical photon) close to the ionization limit and de-excited either to one of the levels $n=8$, 9, or 11, or  to a near high Rydberg state before another de-excitation to one of these levels.
For instance, for Raman scattering of photons close to $H\alpha$ by hydrogen already on level $n=2$, the 912~\AA\ term in Equation~\ref{eq:duv} needs to be replaced by 6560~\AA, which increases the $\Delta_f$ values by a factor of 50.
For UIBs~7.7, $\Delta_f\,\simeq\,50$~\AA.
For NGC7027 the available source power in a 50~\AA\ interval close to $H\alpha$ is $5\,10^{-2}$~W/m$^2$, or  $4\,10^{-3}$~W/m$^2$/sr, which  easily  accounts for the  $2.6\,10^{-4}$~W/m$^2$/sr in-band power (of UIB~7.7) in Beintema et al.'s observations.
\begin{figure}[h]
\resizebox{1.\columnwidth}{!}{\includegraphics{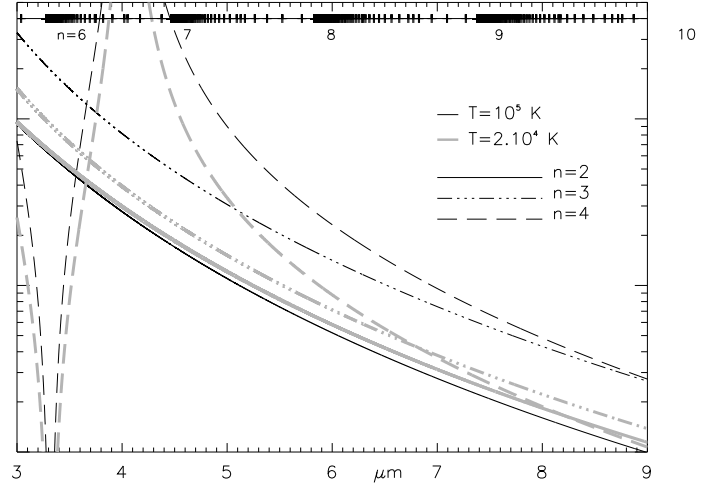}} 
\caption{The Figure plots formula~\ref{eq:low} for two color temperatures of the incident radiation field, $T=10^5$~K (dark lines) and  $T=2\,10^4$~K (thick grey lines).
For each temperature the curves corresponding to level $n=2$,  $n=3$, and $n=4$ are in solid, dash-dotted, and dashed lines respectively.
These curves are scaled in such a way that the curves for level $n=2$  superimpose for the two temperatures.
Raman scattering for level $n=4$ goes through a minimum between $Br\beta$ and $Br\alpha$ and is maximum at $Br\alpha$.
High temperatures neatly favor the contributions from upper hydrogen levels.
For temperatures under $10^4$~K the $n=3$ and $n=4$ curves would be under the $n=2$ line.
This may explain the strong proportion of continuum excess in the spectrum of PNe.
Curves for Raman scattering  with hydrogen left on levels $n=5$ to 11 are not represented but should also appear in this wavelength range.
The additional horizontal axis on the top represents the position of all hydrogen recombination lines in the 3--9~$\mu$m range; near continua correspond to de-excitation from high Rydberg states to levels $n=6$ to 9, as  indicated under the axis. 
Y-axis is logarithmic. 
} 
\label{fig:cont}
\end{figure}
\subsection{Continuum excesses} \label{uibc}
 The behavior of the continuum  underlying UIBs is similar to that of UIB~3.3 (see Section~\ref{uibl}) with respect to the color temperature of the radiation field.
The continuum is particularly strong in PNe such as  NGC7027,  where it dominates over the bands \citep{zavagno92,peeters11}, whereas it is almost imperceptible in lower color temperature radiation fields \citep{lemke98,onaka00,kahanpaa03}.

While UIBs are subordinated to excitation of hydrogen close to the ionization limit,  their underlying continuum could be related to Raman scattering of photons from the continuum that would  bring hydrogen to less energetic levels.
For hydrogen initially at rest and left at a given level $n$, these photons have a wavelength between $Ly_n$ and $Ly_{n+p}$, with $n+p$ less than about 30.
Figure~\ref{fig:cont} shows that these photons  give a continuum expanding towards the infrared (photons between $Ly_n$ and $Ly_{n+1}$) and a series of individual crenels (Section~\ref{patn}) for $p>1$.
The figure also demonstrates that, with respect to the $n=2$ level,  these continua increase steeply with the color temperature of the radiation field, thus justifying the aforementioned behavior of infrared excesses in the  continuum.
\section{NGC7027 and  IC63} \label{ex}
\subsection{NGC7027} \label{ngc7027}
Observations of NGC7027 revealed the existence of UIBs and infrared excesses \citep{gillett67}.
This was also the first PN in which  Raman scattered lines were detected \citep{pequignot97}; further it  exhibits DIBs and ERE \citep{furton90, lebertre93}.
Its bright core is comparable in size with the  Infrared Space Satellite Telescope SWS small aperture (14" x 20"), although the nebula extends arcminutes further \citep[][Figure~3]{phillips07}.
The average hydrogen column density in the nebula is $6\,10^{21}$~cm$^{-2}$ \citep{kastner01}.

A weak, wide dip between 4415 and 4435~\AA\ in the  3300--9200~\AA\ spectrum of NGC7027 observed by \cite{zhang05} is  certainly DIB $\lambda$4428.
There are eighteen emission lines that these authors were unable to identify, including one centered at 6205.73~\AA.
This line likely corresponds to the  6204.5~\AA\ emission line observed by \cite{vanwinckel02} in the Red Rectangle and is bordered  by DIB~$\lambda$6203 on its blue side, which is detected in  Zhang et al.'s spectrum.
The line   is also close in position to DIB~$\lambda$6205.2.
\cite{baluteau95} note   an unexplained absorption at 6613.1~\AA\ in the 6500--10500~\AA\ spectrum of NGC7027 which must be DIB~$\lambda$6613.

The spectrum of NGC7027 thus includes Raman emission and absorption lines.
DIBs in NGC7027 are much weaker than they would appear if observed in the spectrum of a star far behind an interstellar cloud with the same  column density as the nebula  \citep[][]{lebertre93}.
This weakness can be explained in one of two ways.
The absorptions may originate in the nebula, in which case the bands' weakness can be explained by the difference between  scattering at large angles and coherent forward scattering.
Or, as argued by \citet{lebertre93}, DIBs in NGC7027 must be associated with some low column density cloud in the foreground interstellar medium.
\subsection{IC63} \label{ic63}
 \citet{lai20} acknowledge that  "[b]oth DIBs and ERE are produced with highest efficiency in the diffuse ISM, preferring atomic hydrogen environments", and that "DIBs are usually observed in the line of sight towards a distant star, while ERE is observed most easily in a diffuse emission environment adjacent to a hot illuminating star".
In their paper the authors confirm that DIBs  in the spectrum of  a star in the background of a nebula (IC63) have the strength normally expected from the nebula's column density instead of being dimmed, as is systematically the case when reflected nebular light is observed (Section~\ref{ngc7027}).
Notwithstanding the difficulty of finding stars with DIBs in the background of nebulae, there is no doubt that other observations of nebulae and stars in their background  will reveal the same systematic difference between the results of observations in the direction of a star (strong DIBs) versus toward its side (weak or no DIBs).

Lai et al.'s  observations are inconsistent with their conclusion that DIB and ERE carriers  should be close molecules  coexisting in interstellar clouds.
If DIB and ERE carriers are mixed  there is no reason why DIBs should not appear in the same proportion in the spectrum of a background star and  in the reflected light from the nebula. 
As Lai et al. anticipated, the critical issue  is the direction of observation "towards a distant star" versus toward the "environment adjacent to a hot illuminating star".
The angle of scattering --and not specific types of molecules on the line of sight-- determines whether ERE (and UIBs) or DIBs are to be observed.
Consistent with the principles set out  in Section~\ref{scage} of this paper, the findings of  \citet{lai20} highlight the role of the geometrical configuration, thus the importance of the observer's position, in the results of an observation.

\citet{lai20} do not specify the exact molecular composition that DIB and ERE carriers should have, although they seem to privilege "carbonaceous molecules with 18-36 carbon atoms".
It remains hard to understand how these molecules, or the more complex Omont's polyacene  C$_{48}$H$_{26}$ and Peeters et al.'s  C$_{90}$H$_{24}$ or C$_{210}$H$_{36}$, or a "GranPAH" pool of otherwise undefined molecules \citep{omont19,peeters17,andrews15}, could flourish in HI--dominated environments where H$_2$ column densities can be under $10^{16}$~cm$^{-2}$ \citep{meyer82}. 
\section{Conclusion} \label{con}
Just as the ozone layer protects the earth's atmosphere, so too interstellar photodissociation regions (PDRs) shield the inside of interstellar clouds from ultraviolet radiation.
Ultraviolet photons entering PDRs are trapped  and degraded to  lower energy photons for which the interstellar medium is largely transparent.
Light escaping PDRs reaches the observer  in the form of  thermal far-infrared re-emission (after absorption by dust) and, at shorter wavelengths, as broad optical, near and mid-infrared unidentified  "emission" bands, ERE and UIBs (and continuum excesses).

Early hints of a relationship between UIBs and hydrogen transitions were dismissed on the basis of predictions from recombination theory (Section~\ref{uibl}), leaving no  choice but to assign the bands to some complex molecular carrier, the equivalent of ozone for earth's atmosphere.
But this assumption, which led Puget and his group to consider PAHs as a third component of interstellar matter,   does not comport with the primitive  chemistry and strong ultraviolet illumination that must prevail in PDRs.

This paper argued that  atomic hydrogen, the main  component of PDRs, is responsible for ERE, UIBs, and infrared excesses through Raman scattering.
Large Raman shifts associated with hydrogen Raman scattering induce  a transfer of energy from ultraviolet to optical and infrared wavelengths. 
In this sense hydrogen fills the role hitherto attributed to complex organic molecules such as PAHs, but through a different, perhaps more elementary mechanism.
In PDRs ultraviolet starlight is extinguished  by dust (to a lesser extent, close to hydrogen transitions,  by gas).
Trapped ultraviolet photons experience multiple scattering with dust and hydrogen atoms until they are either absorbed by dust (resulting in far-infrared thermal emission) or are Raman scattered by hydrogen (ERE, UIBs, and continuum excesses), in  proportions that remain to be determined.

Hydrogen Raman scattering was  identified shortly after the consensual adoption of the PAH hypothesis, but in such restricted environments (symbiotic stars, PNe) that a connection to ERE or UIBs would have seemed irrelevant.
The grip that PAH theory came to hold on astro-chemistry, along with the narrow focus of HI Raman studies (Raman scattering at optical wavelengths from dust-free gas under strong and high temperature radiation fields) further inhibited recognition of the relevance of Raman theory to  the physics of the interstellar medium.
Raman scattering was nevertheless progressively detected  in most types of objects where ERE and UIBs have been observed: planetary nebulae such as NGC7027, galaxies, and recently the Orion bar.

Attention to Raman scattering in HI media has also been hampered by the widespread belief, including by scholars working in the field, that this is a second order phenomenon.
Raman cross-sections at  source ultraviolet wavelengths are generally a fraction of Rayleigh cross-sections (Figure~\ref{fig:fig2}), but the large Raman shifts associated with Raman scattering by  atomic hydrogen and strong ultraviolet extinctions compensate for the apparent dominance of Rayleigh over Raman scattering.
A comparison of Raman and Rayleigh efficiencies must be done at the emerging optical/infrared wavelengths and include the spectral energy distribution of the radiation field (Section~\ref{ramray}).
For radiation fields with color temperatures larger than $\sim10^4$~K Raman will overwhelm Rayleigh scattering by several orders of magnitude.

The results already obtained from Raman studies and used in this paper are informative on the salient features that can be expected from  hydrogen scattering in the optical  and infrared wavelength range.
An observed spectrum of light Raman scattered by hydrogen is expected to reach ceilings  (because all source ultraviolet photons are Raman scattered) in the vicinity of optical and infrared hydrogen resonances over a spectral extent that depends on the ultraviolet optical depth of the scattering medium \citep{chang15}.
The overall Raman spectrum should thus have a crenel-like structure  (Section 3.5) centered on hydrogen transitions.

However,  in Chang et al.'s dust-free model the wide ERE bump  would require  HI column densities of the order of  $10^{23}$~cm$^{-2}$ which are manifestly too high for the nebulae in which ERE is observed.
Dust  extinction must compensate  for part of the  gas extinction,  allowing for the observation of ERE at substantially lower HI column densities. 
Dust extinction likewise accounts for the finding that UIBs are observed  in dusty HI clouds (Section~\ref{uibext}).

Two findings should eliminate any doubt concerning the involvement of hydrogen Raman scattering in  the physics of  PDRs: the overlap of the ERE/UIB spectral imprint with the spectrum of hydrogen (Section~\ref{uibl}) and the recognition, at their expected position and in the exact proportion, of two pairs of Raman scattered oxygen lines  in DIB catalogs (Section~\ref{dibid}).
This latter finding confirms the distinction that was made in Section~\ref{scage} between Raman scattered emission bands created prior to the scattering (usually in the HII region adjacent to the scattering medium), and Raman absorption lines that  attest to absorptions in the source radiation field before scattering.
In the spectrum of a star's Raman scattered direct light all lines appear in absorption and at a wavelength that is different from the source wavelength. Thus, the reason why  DIBs have been so difficult to identify is that they were not searched for at the right wavelength.
Raman scattering, which broadens emission and absorption lines, also appears as a likely justification for the well-known abnormally large DIB widths.

DIBs confirm  the existence of a scattered light component in the spectrum of stars observed far behind an interstellar cloud.
This component is also the only possible justification of the two-parameter dependency of ultraviolet extinction curves \citep[Section~\ref{scage} and][]{an17,param}.
The difference that the direction of observation (whether toward or on the side of a star) makes in researching DIBs and 2200~\AA\ bumps versus ERE and UIBs highlights the potential variability introduced by geometry in conditioning the results of an observation.
The respective positions in space of the source of light, the scattering medium, and the observer determines whether ERE and UIBs, DIBs, or $1/\lambda^p$ ($p\sim 1$) dust scattering will be observed, with significantly different spectra and polarization in each case (Section~\ref{scage}). 

Since the phenomena discussed thus far are observed at wavelengths longer than 5800~\AA, the question arises of what happens at shorter wavelengths?
Source ultraviolet photons with wavelengths under $Ly\beta$ bring H atoms to metastable states that can de-excite to level $n=2$ with outgoing photons in the vicinity of Balmer transitions short-ward of $H\alpha$, but they can also de-excite to higher levels of hydrogen (Figure~\ref{fig:fig1}).
A diminution of the brightness of  nebulae in the blue compared to the brightness of ERE can be anticipated  because of the spread of Raman scattered photons in different parts of the optical and infrared spectrum (Section~\ref{ramred}).
The degree of diminution can be estimated from the variation of  $n=2$'s branching ratio (see Equation~\ref{eq:max}), which diminishes from the vicinity of $Ly\beta$ (photons Raman scattered around $H\alpha$) to the vicinity of $Ly\gamma$  (photons Raman scattered around $H\beta$) by a factor 0.55 (the decrease slows down over the next Lyman transitions, Figure~\ref{fig:fig1}).
The Raman brightness of a nebula blue-ward  of the ERE bump should therefore be considerably reduced.
Nevertheless, it may  have been detected by  \citet[][see their Figure~4]{vijh05} in the vicinity of the Balmer limit $H_\infty$ (3647~\AA).

Below the Balmer limit,  scattered light by gas is more likely to be of Rayleigh rather than of Raman type, because ionizing photons tend to disappear in PDRs and their Raman cross-section is expected to diminish \citep{marinescu93}.
The slope of the far-ultraviolet rise of ultraviolet extinction curves will be steeper (see Figure~\ref{fig:ray}) than the approximate $1/\lambda^4$ law I mistakenly adopted in previous papers.
The 2200~\AA\ bump could correspond to a transition between Raman and Rayleigh coherent scatterings, whereas the far-ultraviolet rise of extinction curves must be entirely coherent Rayleigh scattered light.

To conclude, the whole extinction curve, including the singular DIB, ERE, and UIB features, can be understood with  a remarkably limited number of assumptions.
These assumptions suffice to enable a synthesis that PAH theorists have sought and never been able to achieve.
There is no need to introduce more constituents into interstellar space than gas and dust, whose existence is established beyond doubt.
Their interaction with light in interstellar space can be analyzed in terms of known physical processes, such as coherent scattering in the forward direction, Raman scattering, or thermal emission.
Optics, atomic physics, and quantum mechanics will supply the complementary theoretical frameworks needed to quantify in a better way than I am able to do the exact form of extinction curves, specify the origin of the strong DIBs, and establish the exact process involved in the production of UIBs. 

\bibliographystyle{model3-num-names}
{}
\appendix{}
\section{K.~Sellgren's hypothesis on near-infrared excesses in nebulae} \label{sell}
In her 1983 Ph.D. thesis, subsequently developed in   \cite{sellgren92}, K.~Sellgren argued that  scattering cannot account  for near-infrared excesses.
According to the paper, optical light scattered from a nebula such as NGC7023  is  Rayleigh scattering by a population of small interstellar particles.
To investigate whether the near-infrared brightness could also be attributed to (Rayleigh) scattering by the same particles,  the authors compared variations of the polarization at blue wavelengths given by  \citet[][]{elvius66}\footnote{There seems to be a confusion in \citet{sellgren92} between \citet{elvius67} and  \citet{elvius66}.} with their own near-infrared observations of the same nebulae.
Since the observed optical polarization was observed to grow and  the infrared polarization to decrease  with wavelength, they concluded that Rayleigh scattering could only account for a small fraction of the near-infrared  brightness of the nebulae, which  therefore had to be produced by  thermal emission from the hypothesized particles (because optical and near-infrared brightnesses  correspond spatially).

As far as I know, no  $1/\lambda^4$ Rayleigh dependency over the full optical spectrum that would attest to the presence of small particles has ever been reported  in the spectrum of nebulae.
Spectra of nebulae between 4000 and 9000~\AA\  consist in a steeply decreasing continuum on which the ERE can sometimes be perceived \citep[see for instance][and also the K.B.~Kwitter \& R.B.C~Henry Gallery of Planetary Nebula Spectra\footnote{Website: web.williams.edu/Astronomy/research/PN/nebulae/}]{schmidt80,perrin92,fd}.
The decreasing  continuum corresponds to the symmetrical halo around bright objects (here the illuminating star) initially studied by \cite{devau58} and   \cite{king71}.
The halo decreases with inverse wavelength as $1/\lambda^p$  ($p\sim 1$), which can either be the well-known extinction law of atmospheric aerosols, or reflected light on the surface of telescopes  \citep{kenknight84,sol1,rr}.
The only optical component of the spectrum of a nebula such as NGC7023 that belongs to the nebula proper is the ERE bump  \citep[][]{fd}.
The  \cite{sellgren92} modeling of the reflected optical light by nebulae as scattered starlight by small grains is thus not supported by observation.

The polarimetry  observations of nebulae (including NGC7023) by \citet{elvius66} are difficult to interpret.
Contamination by de Vaucouleurs' halo of the illuminating stars  (HD200775 for NGC7023) is clearly a problem\footnote{Elvius \& Hall were  aware of this question which they discussed at length. For instance, they wrote for the Merope nebula
 ".. the degree of polarization found for regions five minutes of arc from Merope is not very accurate because of the strong influence of light from this bright star scattered by the atmosphere and instrument." \citep[][p. 260]{elvius66}.} in the blue filters (centered at 3760, 4460, and 5740~\AA) of their observations \citep[][]{fd}.
 \citet[][p. 261, see also references therein]{elvius66} reported that the observed  wavelength dependence  of the polarization in nebulae strongly  resembled  that  of  transmitted (direct) light from the illuminating stars (due to the stars' extinction by foreground interstellar dust,  Section~\ref{pol}).
 In such a case the polarimetry of the  stars and the nebulae  should follow a Serkoski law, increasing from blue to red and decreasing after a maximum to under 8000~\AA, rather than  continuously increasing as supposed by  \citet{sellgren92}.

Further, in the near-infrared part of the spectrum contamination of reflected light from nebulae by the illuminating stars and dust extinction are greatly attenuated.
The near-infrared observations of nebulae should therefore reflect their true polarization. 
 \citet{sellgren84b} noted that the observed polarization angles at near-infrared wavelengths in NGC7023 were "perpendicular to the line between each nebular position and the star which illuminates the visual reflection nebulosity".
As  mentioned above \citet{sellgren92} also  reported that the near-infrared polarization in nebulae was  observed to decrease with increasing wavelength.  
These two findings are characteristic of gas scattering (Raman rather than Rayleigh, Section~\ref{ramred}).
They contravene the conclusion of Sellgren's dissertation that "[t]he primary discovery of the observations described in this thesis is that the near infrared emission detected in the three reflection nebulae NGC~7023, 2023, and 2068 is not due to reflected light"  \citep[][]{sellgren83t}.
In short neither Sellgren's modeling of scattered starlight by nebulae nor their observed infrared polarization can be used to argue that the near-infrared brightness of nebulae is not scattered starlight.

\end{document}